\documentclass[onecolumn]{aa} 
\usepackage{graphicx}
\usepackage{txfonts}
\begin{document}

   \title{Towards a Phylogenetic Analysis of Galaxy Evolution : \\
          a Case Study with the Dwarf Galaxies of the Local Group}

\titlerunning{Towards a Phylogenetic Analysis of Galaxy Evolution}

\author{D. Fraix-Burnet \inst{1}
          \and
        P. Choler \inst{2} 
        \and
        E.J.P. Douzery \inst{3}
       }
 \offprints{D. Fraix-Burnet, \email{fraix@obs.ujf-grenoble.fr}}

   \institute{Laboratoire d'Astrophysique de Grenoble, BP 53, F-38041 Grenoble cedex 9, France\\
         \and
Laboratoire d'\'{E}cologie Alpine, BP 53, F-38041 Grenoble cedex 9, France \\
\and
Laboratoire 
de Paléontologie, Phylogénie et Paléobiologie, Institut des Sciences 
de l'\'Evolution de Montpellier, F-34095 Montpellier cedex 5, France.\\
}

   \date{Received: February 28, 2006; accepted: April 25, 2006}

   \abstract
{The Hubble tuning fork diagram has always been the preferred 
scheme for classification of galaxies. It is based on morphology 
only. At the opposite, biologists have long taken into account 
the genealogical relatedness of living entities for classification 
purposes. }
{Assuming branching evolution of galaxies as a 'descent 
with modification', we show here that the concepts and tools 
of phylogenetic systematics widely used in biology can be heuristically 
transposed to the case of galaxies. }
{This approach that we call 
``astrocladistics'' is applied to Dwarf Galaxies of the Local 
Group and provides the first evolutionary tree for real galaxies.} 
{The trees that we present here are sufficiently solid to support 
the existence of a hierarchical organization in the diversity 
of dwarf galaxies of the Local Group. This also shows that these galaxies all derive from a common ancestral kind of objects. We find that \emph{some} kinds of dIrrs are progenitors of \emph{both} dSphs and other kinds of dIrrs. We also identify three evolutionary groups, each one having its own characteristics and own evolution.}
{The present work opens a new way to analyze 
galaxy evolution and a path towards a new systematics of galaxies. Work on other galaxies in the Universe is in progress.}


   \maketitle
%

\section{Introduction}
\label{intro}

Since Hubble (\cite{hubble22}, \cite{hubble26}), classification of galaxies 
relies largely on morphology: spirals (flattened galaxies with 
spiral arms), barred spirals, ellipticals (galaxies with no obvious 
pattern) and irregulars (Kormendy \& Bender~\cite{kormendy}; Roberts \& Haynes~\cite{roberts}). Subdivisions have been created 
since then (de Vaucouleurs~\cite{devauc}; Sandage~\cite{sandage}) in an attempt to transform morphology into a 
more quantitative parameter. The use of a limited number of additional 
characters---radio and X-ray properties, environment, nucleus activity, gas 
content, star formation, colours and others---has led to the recognition 
of numerous classes which are essentially catalogues of objects. 
This traditional approach of classification might look inadequate 
to describe the now acknowledged complexity of galaxies. Multivariate 
analysis of these characters has been once proposed to tackle 
the problem of synthesizing the classificatory information brought 
by independent characters (Whitmore~\cite{whitmore}; Watanabe et al.~\cite{watanabe}). However, these approaches remain 
based on overall similarities estimations.

Indeed, galaxies are complex systems in which several 
physical and chemical processes govern the evolution of their 
basic constituents: stars, gas, dust, molecules and probably 
black holes (e.g. Vilchez et al.~\cite{evolgal1}). Surprisingly, besides Hubble's primer hypotheses 
on the evolution of galaxies leading to the famous Hubble or tuning 
fork diagram (Hubble~\cite{hubble36}), there has been no further attempt to ground classification 
of galaxies on historical/evolutionary relationships. Nowadays, 
the physical and chemical processes are individually often roughly 
understood and sometimes well modelled (e.g. Sauvage et al.~\cite{evolgal2}), but they are not 
collectively integrated, and grasping the causes of galaxy diversity 
in their entire complexity remains a difficult task.

Similar concerns have already been addressed in the study 
of biological diversity. Nearly 150 years ago, Darwin (\cite{darwin}) suggested 
that the hierarchical classification of living organisms should 
reflect their genealogical relationships. This was an incredibly 
successful idea that drove biological systematics into a new 
era. There is now a common agreement that a natural 
classification should be derived from phylogenetic trees, i.e. 
a branching structure describing the evolutionary relationships 
of a set of biological entities or taxa (Wiley et al.~\cite{wiley}). Accordingly, the 
concepts and methods of phylogenetic systematics have been successfully 
applied to other sets of entities for which historical or evolutionary 
relationships could be documented; examples are found in linguistics (Wells~\cite{wells})
and stemmatics (Robinson \& Robert~\cite{robinson}).

We therefore hypothesize that galaxies can be classified 
in a natural hierarchy of nested groups reflecting evolution. 
Our purpose is thus to introduce the parameter ``time'' in the 
classification scheme, using all available and suitable descriptive 
characters, and to propose an integrated way to represent galaxy 
diversity. We named this approach ``astrocladistics'' (Fraix-Burnet et al.~\cite{FCDa}; Fraix-Burnet~\cite{fraix}). 
The method and associated concepts are presented in detail elsewhere together with 
an analysis of two samples of simulated galaxies (Fraix-Burnet et al.~\cite{jc1}, ~\cite{jc2}). In this 
paper, we present the first application of astrocladistics on 
real galaxies. We chose to focus on the Dwarf galaxies of the 
Local Group for which a reasonable amount of complete and homogeneous 
data were available. According to the ``hierarchical scenario''\footnote{This 
is not to be confused with the hierarchical organisation of galaxy diversity 
dealt with in this paper.} of galaxy formation, these galaxies could belong to small dark matter halos that are the building blocks of larger structures that may host big galaxies. 
In this respect, dwarf galaxies are not considered to be formed by the merging of smaller structures, but because of the relatively shallow gravitational well they are certainly subject to disturbances and sweeping. However, this hierarchical scenario of galaxy formation has some difficulties, like predicting too many small structures which are not seen in the form of dwarf galaxies (e.g. Moore et al.~\cite{moore}, Venn et al.~\cite{venn}). The 
Local Group could also represent an evolutionary microcosm where 
the environments have been somewhat similar for all the dwarfs. 
Here we scored a variety of characters, and inferred the most 
parsimonious history that describes the evolution of Dwarf Galaxies 
of the Local Group.

   \begin{table}
	\centering
      \caption[]{List of characters used in the cladistics analysis. The 
constraint on the evolution of each character is indicated by 
``o'' for ``ordered'' and ``u'' for ``unordered''.}
         \label{charlist} 
         \begin{tabular}{rlc}
            \hline
            \noalign{\smallskip}
           & Character      &  Constraint  \\
            \noalign{\smallskip}
            \hline
            \noalign{\smallskip}
1   & Ellipticity                               & u    \\
2   & Core radius                               & o    \\
3   & V Luminosity (integrated)                 & o    \\
4   & B-V (integrated)                          & o    \\
5   & U-B (integrated)                          & o    \\
6   & Total mass                                & o    \\
7   & Central mass density                      & u    \\
8   & HI mass                                   & o    \\
9   & Mass / luminosity ratio                   & o    \\
10  & HI mass / total mass ratio                & o    \\
11  & HI mass / B luminosity ratio              & o    \\
12  & Dust mass                                 & o    \\
13  & HI flux                                   & o    \\
14  & CO flux                                   & u    \\ 
15  & H$\alpha$ flux                            & o    \\ 
16  & Fe / H ratio                              & o    \\ 
17  & O / H ratio                               & u    \\ 
18  & N / O ratio                               & o    \\ 
19  & Star Formation Rate                       & u    \\ 
20  & Rotational velocity                       & u    \\ 
21  & Velocity dispersion of ISM                & u    \\ 
22  & Rotational velocity / Velocity dispersion & o    \\ 
23  & Maximum rotational velocity               & u    \\ 
24  & Central velocity dispersion               & u    \\
            \noalign{\smallskip}
            \hline
         \end{tabular}
   \end{table}
   
   \begin{figure*}
   \centering
   \includegraphics[width=15 true cm]{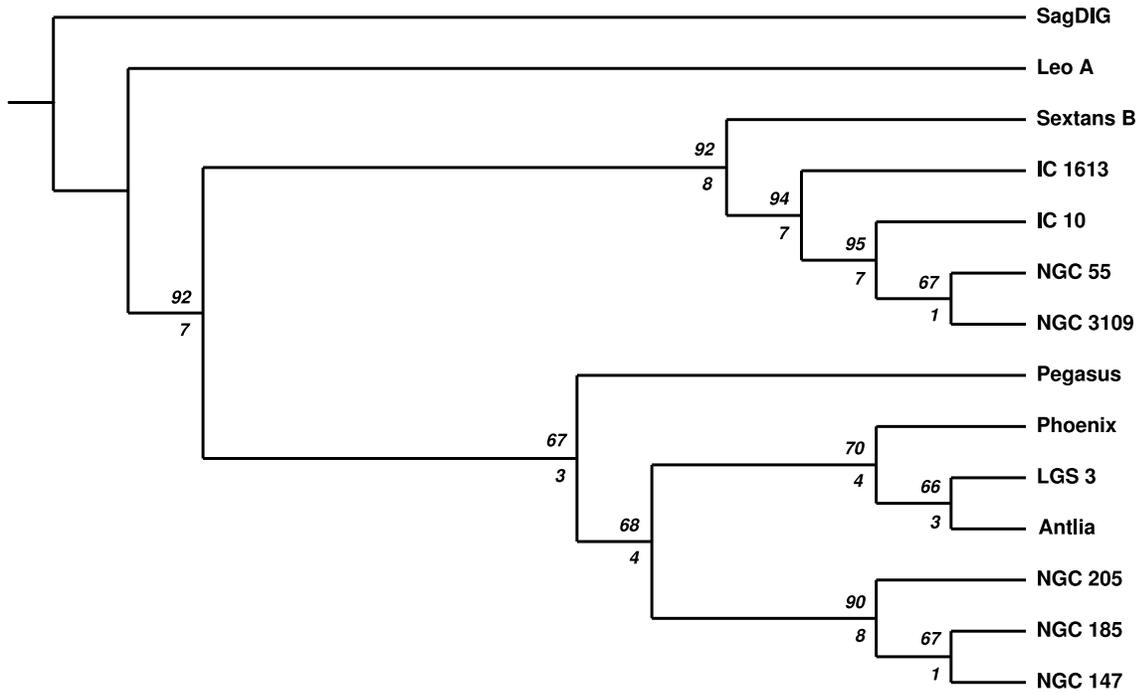}
   \caption{Tree of 14 galaxies obtained by a maximum parsimony analysis 
of 24 characters as described in the text. It has 167 steps, Consistency Index=0.68, Retention Index=0.75, Rescaled Consistency Index=0.51. Numbers at each node are bootstrap (above) and decay (below) values. Bootstrap percentages are obtained after 1,000 resampling of characters. Decay indexes correspond to the minimum number of character-state changes to be added to tree length to break the corresponding node.}
              \label{figure14}%
    \end{figure*}

   \begin{figure*}
   \centering
   \includegraphics[width=17 true cm]{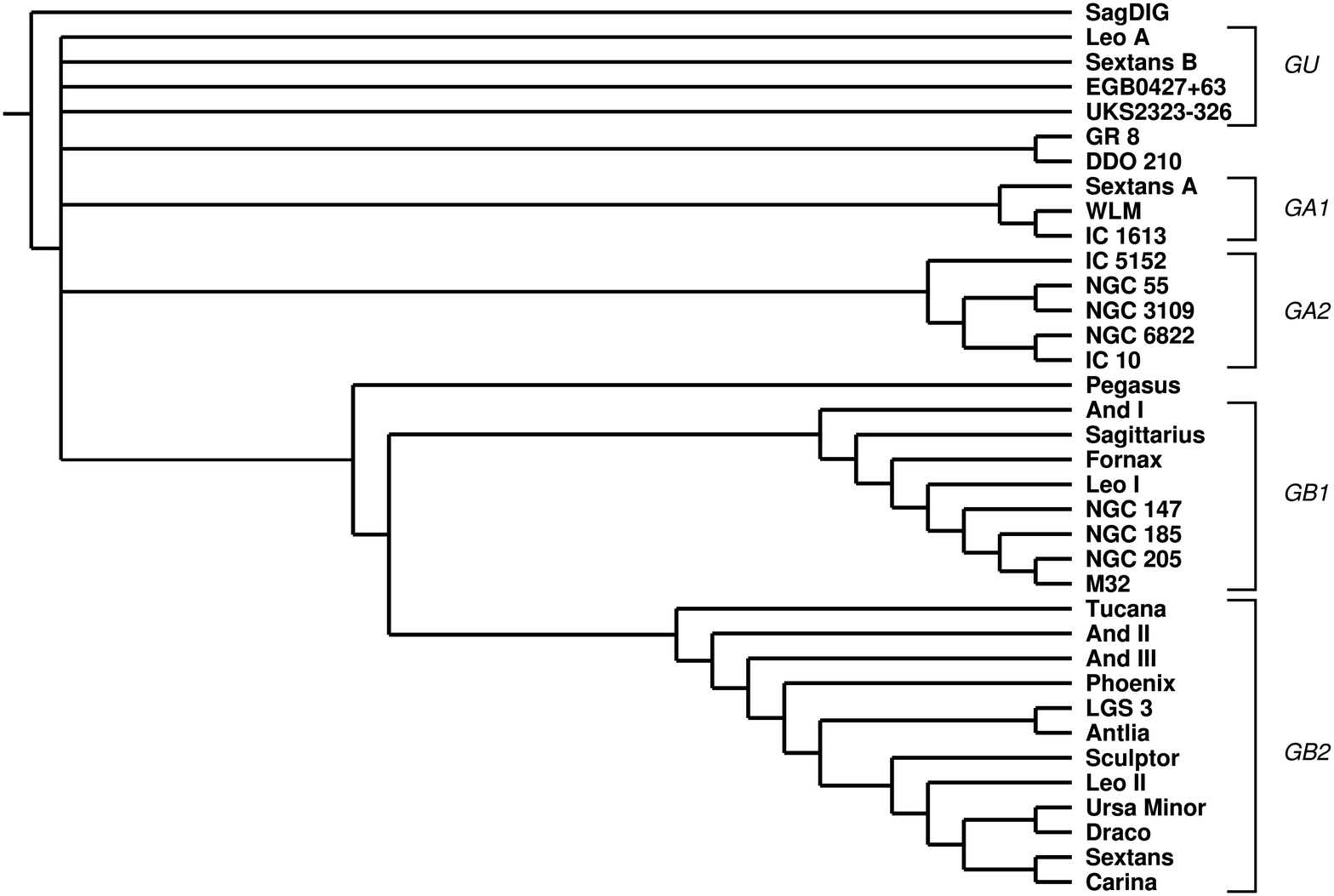}
   \caption{This tree with the 36 Dwarf galaxies of the Local Group, is the strict consensus of 1041 most parsimonious trees having 317 steps each. Groups corresponding to sub-structures are indicated and discussed in the text.}
              \label{figuretot}%
    \end{figure*}

\section{Methods}

\subsection{Conceptual framework}

A cornerstone of phylogenetic 
systematics is to focus on derived character states to infer 
common ancestry relationships (Hennig~\cite{hennig}). Such characters may be viewed 
as evolutionary novelties appearing in a particular lineage. 
It is assumed that two closely relative objects (also called taxa because they can be individuals, groups or species) share derived characters, 
which presumably originate in their common ancestor. Regarding 
classification, phylogenetic systematics defines clades (evolutionary groups) as groupings 
that are significant with respect to evolution in the sense that 
they contain all offspring from a common ancestor, and their 
members share common derived states, that is to say a common 
history.

Now, let's consider two identical galaxies hence belonging 
to the same class. If they evolve independently and isolated, 
the number of processes at work within each galaxy makes it probable 
that after a given time, they will be different enough to be 
put into two different classes. For instance, if they were both 
of spiral shape, they will remain as such but one could have 
developed a bar or had a starburst activity modifying its stellar 
population. In addition, galaxies live in an environment made 
of intergalactic gas and gravitational potential fashioned by 
dark matter and other galaxies (Sauvage et al.~\cite{evolgal2}). If two galaxies interact with 
each other, the dynamics will most of the time be affected differently 
in both because of asymmetries in the encounter, so that they 
could then be put in two separate classes. In all cases, one 
class gives birth to at least two classes. This is the definition 
of branching evolution. Its driver is mainly the randomness of 
external interactions, and to a lesser extent that of internal 
processes. This branching evolution results in a hierarchical or tree-like organisation of the diversity.

All details on astrocladistics are given in Fraix-Burnet et al.~\cite{jc1, jc2}.

\subsection{Data}
\label{data}

We used data from Mateo (\cite{mateo}) on 36 galaxies 
and identified 25 characters. For each character, we discretized 
the whole range of values in up to 8 bins plus the unknown value 
``?''\footnote{Tables are available on http://hal.ccsd.cnrs/aut/fraix-burnet}. These are assumed to be evolutionary 
states. For instance, if we accept that metallicity increases 
with time, then a very low value would be an ancestral state 
whereas a high value would be a derived state. This has to be adapted to the sample under study.
We imposed 15 characters to be ordered (see Table~\ref{charlist}), i.e. changes between two adjacent states are more probable 
than between distant ones, independently from the sense of the change. The morphological character (dIrr, Intermediate and dSph)
was discarded in the analysis because it notably differs from 
the other ones in being the only qualitative property and somewhat 
subjectively defined. It was coded into three states merely for projection onto the result trees. 
We thus ended up with a matrix of 36 objects described by 24 characters, the total percentage of unknown values being 33\%).

The outgroup, used to root the tree ---i.e., to orientate the evolutionary 
processes--- indicates the ancestral states of the characters for which transformation 
can be hypothesized within a reasonably accepted paradigm. Given the complicated evolutionary behaviour of Dwarf galaxies, for this first astrocladistics analysis, we decided to choose a galaxy of the sample. The advantage is that the result tree can be easily rerooted with any of the sample objects depending on one's own guess, because the structure of the tree does not change. SagDIG is chosen here as root because it has the highest ratio MHI/M of the sample. It is interesting to note that SagDIG seems to be at the verge of our Local Group (van den Bergh~\cite{bergh2000}).

\subsection{Tree construction}

An important aspect of phylogenetic 
systematics is to avoid the grouping of objects based on similarities due to evolutionary convergences or reversals. 
In phylogenetic analysis, maximum parsimony (Fitch~\cite{fitch}) is one optimality 
criterion for selecting among competing trees. It states that the most probable evolutionary scenario is the simplest one as measured by the total number of character state changes (called steps). The most parsimonious tree thus corresponds to the simplest evolution scenario compatible with all the input data.
The use of the parsimony principle also minimises the number of convergences or reversals. We used maximum parsimony to identify 
the tree best representing our data matrix. Since there is an important number of unknown values in the data, the full tree obtained with the 36 galaxies is not extremely robust in the sense that slightly different trees could also fit the data. To assess its significance, we thus looked for smaller trees and kept the more robust one we can find with a subsample as large as possible. The evaluation of the 
reliability of tree nodes was made through bootstrap (percentages of node occurence obtained after 1,000 resampling of characters, Felsenstein~\cite{felsenstein}) and decay (minimum number of character-state changes to be added to tree length to break the corresponding node, Bremer~\cite{bremer}) estimates. All calculations were conducted with PAUP*4.0b10 (Swofford~\cite{paup}) and the tree interpretation was done with Mesquite (Maddison \& Maddison~\cite{mesquite}).

\section{Results}
\label{results}

We found a fully resolved tree with 14 galaxies (Fig.~\ref{figure14}). It has a total of 167 steps (character state changes), bootstrap and decay values are high for nearly all nodes (Fig.~\ref{figure14}), indicating that most tree nodes are strongly supported by the data. No better trees could be found with 14 or more objects. Regarding the groupings and evolutionary behaviours of characters, that is the resulting phylogeny, all robust trees we found with 14 or slightly less galaxies are all compatible, as they are with the full tree discussed below.

Performing an analysis with the full sample of 36 galaxies, 1041 most parsimonious trees were found, each having 317 steps. The strict consensus of all of them is shown in Fig.~\ref{figuretot}. It is remarkably well resolved despite the important number of unknown values. The strict consensus means that the nodes present on the tree are found in all the 1041 individual trees. We must warn that this tree is not as robust regarding bootstrap and decay estimators as the previous one so that its very detailed structure should be taken with some caution. However, the two trees are fully compatible and the character behaviours, as will be discussed below, are identical, pointing toward the same evolutionary history of diversification.

Generally speaking, a galaxy situated far from the root (SagDIG) is more differentiated from it. It is not necessarily more evolved since it depends on the timescale of each evolutionary process (see \cite{jc1, jc2} for a more complete discussion). In this paper, for simplification at the level of the discussion and because we lack an evolutionary clock, we will identify diversification and evolution.

The very existence of resolved and robust trees implies that all these galaxies share a common ancestor with respect to SagDIG, that is they all derive from the same kind of objects. In addition, diversity in the Dwarf galaxies of the Local Group arranges itself in a hierarchical way.

Five groups are identified in Fig.~\ref{figuretot} and defined in Table~\ref{clades}. These groups correspond to structures on the tree, mainly lineages, except \textit{GU} which gathers a set of unresolved branches for convenience in the discussion. After SagDIG, the root of the tree, there are four groups with unresolved respective phylogenies, \textit{GU, GA1, GA2} and \textit{GB} (\textit{GB1} and \textit{GB2}) plus the pair GR~8--DDO~210. In contrast, the branch after Pegasus diverges in two clear subtrees, \textit{GB1} and \textit{GB2}, which are by the way the most diversified from SagDIG among the whole sample. There are three pairs within \textit{GB2}: LGS~3--Antlia, Ursa~Minor--Draco and Sextans--Carina. On the small tree (Fig.~\ref{figure14}), the phylogeny is fully resolved, and the above groups are present with at least one object representative of each (indicated in Table~\ref{clades}). After the branch toward Leo~A, there are only two diverging branches, one leading to \textit{GA1} then \textit{GA2}, the other one finally leading to another binary bifurcation toward \textit{GB1} and \textit{GB2}. Note that \textit{GA} (1 and 2) and \textit{GB2} have very high bootstrap and decay values, indicating that they probably constitute true evolutionary group. The support for \textit{GB1} is somewhat less but still very significant. In other words, the small tree of Fig.~\ref{figure14} is a phylogeny of the groups defined in Table~\ref{clades}.

The reasons for the tree structure lies in the evolution of the characters. Their values are projected onto the larger tree in Fig.~\ref{figureproj} in order to visualize both the character evolution and the group properties. The latter are summarized in Table~\ref{clades}. It appears that, except for \textit{GU} which by construction cannot be considered to form an evolutionary group because the relative placement of galaxies in \textit{GU} is unknown, groups can be unambiguously characterized by a set of properties. The three groups \textit{GA2, GB1} and \textit{GB2} have unique characteristics. Group \textit{GA1} resembles \textit{GA2} but lacks a high V luminosity. Following the small tree in Fig.~\ref{figure14}, these two groups are related, implying an increase in V luminosity with evolution along this branch. It is difficult to conclude with the present data whether they are one or two different lineages.

On the contrary, there are undoubtedly two species identified by the two groups \textit{GB1} and \textit{GB2}. In the first case, galaxies are specifically bright in V, have a high central mass density but a low central ratio M/L, together with a low MHI/Blum. In the second case, galaxies are of low V luminosity with a high and increasing (along this branch) central ratio M/L, a decreasing MHI/Blum, and may be more importantly a very low metallicity Fe/H.

Assuming that SagDIG, as the root of the tree, resembles the common ancestor to Dwarf galaxies of the Local Group, a few characters seem to evolve regularly with the evolutionary scenario revealed by our tree: colours (B-V and U-B) might globally increase, while MHI/M, the HI mass, MHI/BLum might decrease. But the structure of the tree is not very regular, and most characters, like the V luminosity, the total mass, or the dust mass, have different behaviours depending on the branches. Hence, the tree structure is explained by multiple parameters together with possibly a very few global evolutionary trends. The seemingly complex history of diversity in this sample might be due to the particular nature of dwarf galaxies, and this shows that the different evolutionary stages cannot be established with only one or two descriptors.

This statement could seem to be contradicted by the morphology behaviour as can be seen on the corresponding projection in Fig.~\ref{figureproj}. Despite this character was not used to build the tree, a perfect dichotomy appears between galaxies before Pegasus (dIrrs with groups \textit{GU}, \textit{GA1} and \textit{GA2}) and after it (dSphs with groups \textit{GB1} and \textit{GB2}).
This galaxy is of intermediate type. Interestingly, our 14-galaxy tree (Fig.~\ref{figure14}) diverges after Pegasus into two groups, one of which is composed exclusively with three other intermediate-class galaxies (Antlia, LGS~3, Phoenix). But it corresponds to \textit{GB2} (Fig~\ref{figuretot}) which also have dSphs. The last intermediate-type galaxy (DDO~210) is paired with GR~8 among the dIrrs. As clearly demonstrated by the structure of the tree, the dichotomy dSph-dIrr is far too simplistic to represent the diversity of dwarf galaxies, even solely in the Local Group, and misses all groups identified in this paper that are based on the information included in 24 characters. Indeed, 
spheroidal galaxies are all gathered in the group \textit{GB}. These galaxies constitute a separate evolutionary lineages, and Fig.~\ref{figure14} stipulates more precisely that they derived from ancestors of irregular morphologies.  However, this is not enough to depict the diversity and evolutionary history of these galaxies since there are probably at least two kinds of irregulars and certainly two rather different kinds of spheroidals. The character projections (Fig.~\ref{figureproj}) illustrate the well known fact that the spheroidal dwarf galaxies have lost their HI gas, but goes further in describing with more detail the several kinds of evolutionary groups identified above.

   \begin{table}
	\centering
      \caption[]{Characteristics of groups identified in Fig.~\ref{figuretot}. Starred names indicate galaxies present on the small tree in Fig.~\ref{figure14}. The unique characteristics belong only to the group, while the other characteristic are found in other groups. The words ``increasing'' or ``decreasing'' refer to evolution within the group as oriented on the tree rooted with the SagDIG galaxy. The morphology (in parentheses) was not used to derive the tree.}
         \label{clades} 
         \begin{tabular}{llll}
            \hline
            \noalign{\smallskip}
            Group      &  Members   & Unique characteristics & Other characteristics \\                      
            \noalign{\smallskip}
            \hline
            \noalign{\smallskip}
GU  & Leo A *      &  & High MHI/BLum\\
    & Sextans B *  &  & (dIrr)\\
    & EGB0427+63   &  &\\
    & UKS2323-326  &  &\\
            \noalign{\smallskip}
GA1 & Sextans A    &  & Low Central Mass density\\
    & WLM          &  & Low central M/L\\
    & IC 1613 *    &  & High MHI/BLum\\
    &              &  & (dIrr) \\
            \noalign{\smallskip}
GA2 & IC 5152      & High Mass & High V Luminosity\\
    & NGC 55 *     & High HI Mass & Low Central Mass density\\
    & NGC 3109 *   & High Dust Mass & Low central M/L\\
    & NGC 6822     & High HI Flux & High MHI/BLum\\
    & IC 10 *      & High CO Flux & (dIrr) \\
    &              & High H$\alpha$ Flux & \\
    &              & High SFR & \\
    &              & High Vrot/Vdisp & \\
            \noalign{\smallskip}
GB1 & And I        & High U-B  & Increasing V Luminosity \\
    & Sagittarius  & Highly increasing Central Mass density & Increasing B-V \\
    & Fornax       & Decreasing central M/L & Very low HI Mass \\
    & Leo I        & Low MHI/BLum & Very low MHI/M \\
    & NGC 147 *    &  & (dSph)\\
    & NGC 185 *    &  &\\
    & NGC 205 *    &  &\\
    & M32          &  &\\
            \noalign{\smallskip}
GB2 & Tucana       & Low and decreasing V Luminosity & Increasing B-V \\
    & And II       & Increasing central M/L  & Increasing Central Mass density \\
    & And III      & Decreasing MHI/BLum  &  Very low HI Mass \\
    & Phoenix *    & Low Fe/H  & Very low MHI/M \\
    & LGS 3 *      & & (dSph) \\
    & Antlia *     &  & \\
    & Sculptor     &  &\\
    & Leo II       &  &\\
    & Ursa Minor   &  &\\
    & Draco        &  &\\
    & Sextans      &  &\\
    & Carina       &  &\\
            \noalign{\smallskip}
            \hline
         \end{tabular}
   \end{table}

\section{Discussion}

The trees show a hierarchical organization of the 
Dwarf Galaxies of the Local Group with well identified evolutionary groups. 
This result is particularly remarkable since these galaxies evolve 
in the same microcosm which probably
mean less diversity. But more importantly, the two main hypothesis of astrocladistics, a common ancestor and a hierarchical organisation of diversity, are thus verified. The present result strongly stimulates the use of this approach to other galaxies, as is currently under progress.

The existence of a hierarchy in galaxy diversity makes it possible 
to track ancestral types of galaxies. Our result shows that the Dwarf galaxies of the Local Group very probably have a common ancestor, meaning that they derive from the same kind of objects. This seems to be in contradiction with Ricotti \& Gnedin (\cite{ricotti}) who assume two different initial formation processes for irregulars and spheroidals. Even though SagDIG is supposed here to resemble the common ancestor of our sample, it is impossible to tell how much it resembles the common ancestor of all dwarf galaxies. Study of much more objects is needed. However, our result might not be incompatible with Ricotti \& Gnedin's result since it depends on the definition of both ancestor object and formation concept (see Fraix-Burnet et al.~\cite{jc1, jc2}). In addition, dwarfs are currently thought to form within small dark matter structures but this formation process is far from being understood and even definitively established. 

Assuming different stellar formation histories, Ricotti \& Gnedin define three groups: survivors, true and polluted fossils. Only the true fossils correspond relatively well to one of our groups, \textit{GB2}, except for Carina, LGS~3 and Leo~II that they classify as polluted fossils. But our classification lies on other descriptors than only the stellar population. In the same manner, Grebel (\cite{grebel}) presents a diagram tentatively summarizing the history of dwarf spheroidals based on star formation episodes. If we compare the exemplar galaxies of her categories with our tree, we find that Draco (``old'') and Carina (``episodic'') both belong to \textit{GB2}, Fornax and Phoenix (both ``young'') belong respectively to \textit{GB1} and \textit{GB2}, and Leo~I (``intermediate'') is close to Fornax. Again, the two classifications do not match for the same reason as above, indicating that the star formation history is certainly insufficient to depict the whole diversity of the Dwarf galaxies. On the contrary, our data may be inadequate to characterize the stellar population in detail. Since an astrocladistics analysis reflects the knowledge at a given epoch and is thus never finished, extension of the present work in the future will have to include more descriptors as they become available for most of the galaxies.

Because the cladogram recapitulates evolutionary information of several characters, we think it could give 
a better insight on possible caveats of wordings like ``young galaxy'' and ``old galaxy''. All the galaxies of the sample are contemporaneous, but some resembles more the ancestor than others, that is they look like ``old'' species. They are not necessarily ``old'' objects. Likewise, galaxies at the bottom of the tree are more diversified, but cannot be said ``younger''. For instance, Ursa Minor is often believed
to have little evolved (Mateo~\cite{mateo}; Carrera et al.~\cite{carrera}; Mighell~\cite{mighell}) because it has old stars. At the opposite, IC~10 have had a strong and recent starburst activity (Mateo~\cite{mateo}; Wilcots \& Miller~\cite{wilcots}). From our cladistic analysis, it would be meaningless to qualify one ``younger'' than the other, rather Ursa Minor seems to be more diversified from the ancestor as compared to IC~10. They do belong to two different species, each with its own evolutionary pathway, that diverged sometime in the history of these galaxies and all their progenitors (see Fraix-Burnet et al.~\cite{jc1, jc2}). Hence, we think it is tricky to speak of ``old'' or ``young'' galaxies based on star formation history alone. It is much preferable to refer to the stage of diversification.

The evolution of dwarf galaxies of the Local Group is often reduced to the segregation between dIrrs and dSphs  (see e.g. Mateo~\cite{mateo}). For instance, one explanation for the depletion of HI gas in dSphs could be the proximity of a big galaxy. In Fig~\ref{figureproj}, each galaxy name has a colour corresponding to the sub-group to which it belongs (Mateo~\cite{mateo}). All members of the Milky Way family are spheroidals, but they do not belong to the same evolutionary group. There is no clear correlation with distance to the Milky way. 
For the M31 (Andromeda) family, the four closest satellites (M32, NGC~205, NGC~147 et NGC~185) are grouped together in \textit{GB1} with, among some others, Sagittarius, a very close satellite of our Milky Way. The other members are spread all over the tree with EGC0427+63 and IC~10 being in \textit{GU} and \textit{GA2} respectively. The NGC~3109 family is spread over the tree, while the Local Group Cloud members are spread over \textit{GU, GA1}, \textit{GA2} and even \textit{GB2} for Tucana. Note that SagDIG belongs to this last family, but is certainly the farthest member of the Local Group being at its verge (van den Bergh~\cite{bergh2000}).
Our conclusion is that the measured distance to a big galaxy is not enough to explain the diversity of the Dwarf galaxies of the Local Group even though companions of the two big galaxies of our Local Group nearly all belong to the groups \textit{GB1} and \textit{GB2} together with only three more ``isolated'' objects.
One must keep in mind that the measured distance is the current distance, while the properties of galaxies are the result of a long history along a probably complicated trajectory that saw this distance change considerably.

Another example is given by conclusions like dIrrs being unlikely progenitors of dSphs (e.g. Grebel~\cite{grebel}). This kind of statement ignores the complexity of galaxies and their evolution. Comparing the two morphological types globally cannot be successful due to the large variety of objects within each class (see e.g. Mateo~\cite{mateo}). The tree on Fig~\ref{figuretot} and the character projections (Fig~\ref{figureproj}) illustrates that the evolutionary stage of a galaxy cannot reasonably be assessed with the sole morphology criterion.
Our result rather shows that \emph{some} kinds of dIrrs are progenitors of \emph{both} dSphs and other kinds of dIrrs that diverged and developed their own lineage.

Our choice of SagDIG as root provides a globally satisfactory evolutionary scenario. This choice could easily be changed without modifying the structure of the tree, but then the interpretation would be different.
Regularity of evolution was imposed to some characters (Sect.~\ref{data}, Table~\ref{charlist}), but this constraint does not imply monotonic behaviours as can be seen on the tree (Fig.~\ref{figureproj}, Sect.~\ref{results}). More importantly, characters can evolve differently in different lineages. For instance, the V luminosity regularly increases in \textit{GB1}, while it decreases in \textit{GB2}, making a global evolutionary behaviour quite complex and of poor significance. 

We have identified at least three evolutionary groups (Table~\ref{clades}). The first one (\textit{GA2}) is composed of massive galaxies with a high star formation rate, lot of HI gas and dust, high fluxes of HI, CO and H$\alpha$, and a high ratio between rotational and dispersion velocities. The second one (\textit{GB1}) has relatively red galaxies (high U-B), with an increasing central mass concentration and a decreasing central M/L, while the ratio MHI/B luminosity is low. The third one (\textit{GB2}) has galaxies with low and decreasing V luminosity, an increasing central M/L, a decreasing MHI/B luminosity and a low Fe/H. In summary, the group \textit{GA2} is rather active and massive, \textit{GB1} accretes gas toward its center and increases its global V luminosity, and \textit{GB2} looks like a dead end with gas-poor galaxies with decreasing V luminosity. These are three different evolutionary paths grouping objects with similar histories that models will have to understand.

Since it is based on observational data, the evolutionary scenario depicted by the tree 
is globally consistent with current thoughts about dwarf history and physics (e.g. Mateo~\cite{mateo}; Carrera et al.~\cite{carrera}; Wilcots \& Miller~\cite{wilcots}; Gallagher \& Wyse~\cite{gallagher}). It has the advantage of being built without choosing some particular parameters, and is the unique way to synthesise the information contained in such a multivariate problem. In addition, astrocladistics introduces the evolution in the analysis itself, providing a direct view of the diversification process.

The results are based on the data and naturally depends on their quality, in particular on distance estimations of galaxies. Regarding 
error bars in the original data, it is not expected that they 
bring much noise in the result because initial quantitative data 
are binned (coded). In the same manner, the effect of any aberrant point is smoothed out in the analysis since the result is a synthesis of all the input information.
In any case, a cladistics analysis has this 
invaluable quality of being entirely transparent and falsifiable, 
so that it is always possible to modify original values, modify the way the coding is done, introduce weights 
to characters, and compare the resulting phylogenies.

\section{Conclusion}

We conclude that branching evolution is the dominant diversification process 
among the 36 Dwarf Galaxies of the Local Group since the diversity is organized in a tree-like or hierarchical way. Phylogenetic systematics is thus applicable to astrophysics 
and provides a powerful tool to apprehend galaxy formation and 
evolution. The tree depicts relationships between galaxies 
by taking into account physical and chemical character changes. 
It thus provides the opportunity to formulate new inferences on galaxy evolution by providing a synthetic view of current knowledge. 

We have identified at least three different evolutionary groups, each one having its own characteristics and own evolution: one is composed of massive and star forming galaxies, another one has galaxies that seem to increase the luminosity while transferring mass toward their centre, and the last one looks like the dead-end of dwarf galaxy evolution with dimming and concentrating objects.

Morphology classification appears to be too simplistic to understand the evolution of dwarf galaxies. From the trees we have found, we derive that \emph{some} kinds of dIrrs are progenitors of \emph{both} dSphs and other kinds of dIrrs. Two of our evolutionary groups are composed of dSphs, the other one of dIrrs, the intermediate type being spread over the tree.

We must stress that the phylogeny found is entirely based 
on the data input. The best way to avoid introducing a subjective 
bias is to consider all available characters excluding obviously 
redundant ones, as is done in phylogenetic systematics. As new data 
and/or new characters become available and introduced in the 
analysis, the current phylogeny might change somewhat in the 
future. This is normal process of knowledge progress. In the particular case of the Dwarf galaxies of the Local Group, spectrophotometric observations will certainly be invaluable for astrocladistics, but this would ideally require homogeneous data on a significant sample to be available.

Extension of this study to much more types of galaxies in the Universe 
is in progress and will possibly yield the basis for a new taxonomy of galaxies.

\begin{acknowledgements}
We thank an anonymous referee for suggestions that lead to a large improvement of the paper.
Most computations of the results presented in this paper were 
performed under the CIMENT project in Grenoble. E.J.P.D. has 
been supported by the Genopole Montpellier Languedoc-Roussillon 
and the Action Bioinformatique inter-EPST of 
the CNRS, and this work represents the contribution N\ensuremath{^\circ} 2006-030
of the Institut des Sciences de l'Evolution de Montpellier (UMR 
5554 -- CNRS).
\end{acknowledgements}

\Online{

   \begin{figure*}
   \centering
   \includegraphics[width=16 true cm]{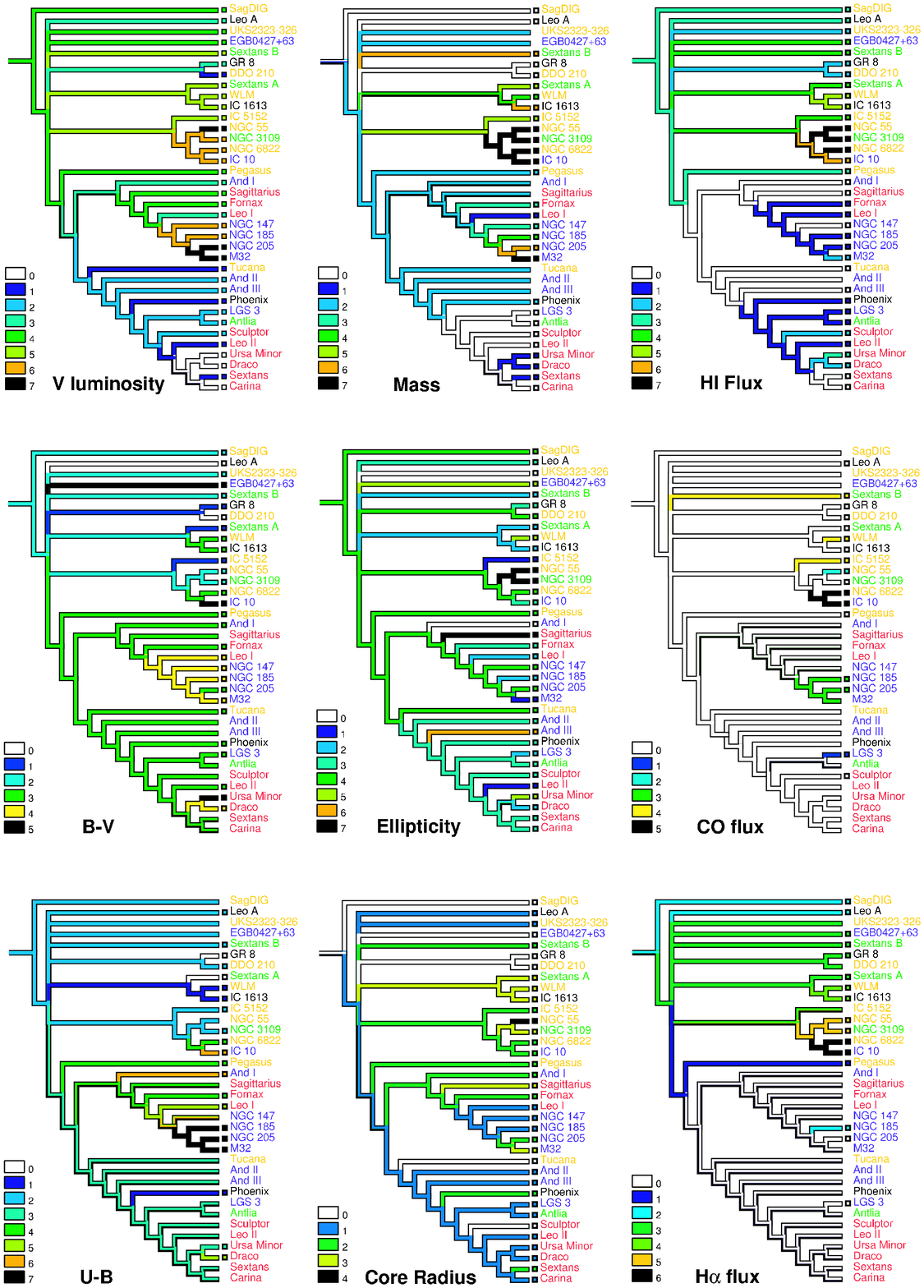}
   \caption{Colour coded projection of characters with the convention that the white code is always the lowest value of the corresponding parameter. Small squares at the leaves (before galaxy name) indicate a documented value for that character. The other leaves are given a value deduced from a parsimony analysis, providing a prediction for the corresponding galaxy. Colour code for galaxy names is given at the end of figure. Morphology is plotted but was not used to derive the tree.}
              \label{figureproj}%
    \end{figure*}

   \begin{figure*}
\begin{center} 
   \includegraphics[width=16 true cm]{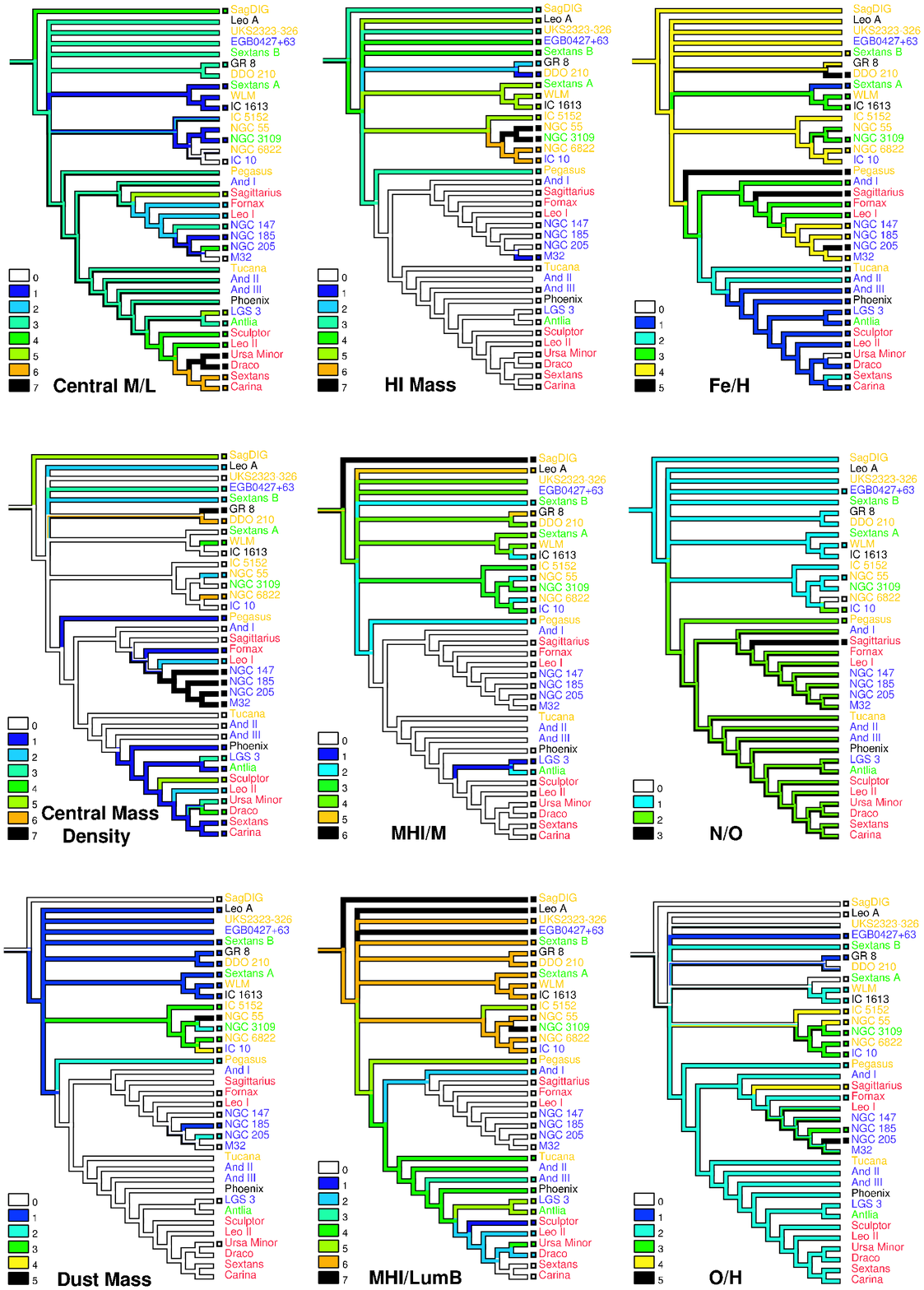}
\end{center}
\textbf{Fig.~\ref{figureproj}.} \textit{(continued)} 
    \end{figure*}

   \begin{figure*}
\begin{center} 
   \includegraphics[width=16 true cm]{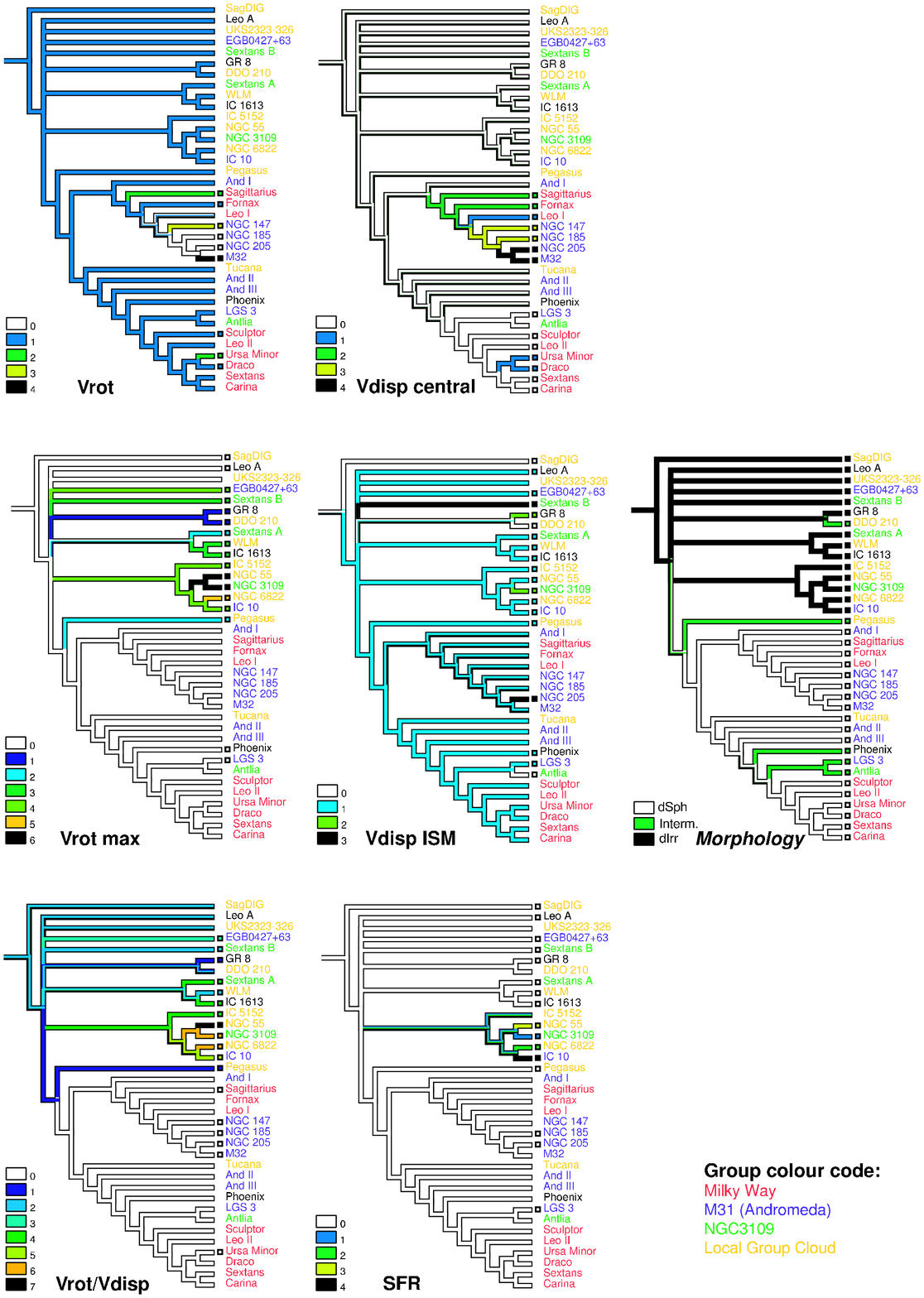}
\end{center}
\textbf{Fig.~\ref{figureproj}.} \textit{(continued)} 
    \end{figure*}
}

\end{document}